\documentclass[aps,prl,twocolumn,groupedaddress,floats,showpacs]{revtex4}
\usepackage{latexsym}
\usepackage{dcolumn}
\usepackage[dvips]{graphicx}
\usepackage{amssymb}
\usepackage{graphics}
\usepackage{amsmath}
\usepackage{epsf}

\begin{document}
\author{E. H. Hwang, S. Adam and S. Das Sarma}
\title{Carrier transport in 2D graphene layers}
\affiliation{Condensed Matter Theory Center, Department of Physics, 
University of Maryland, College Park, MD 20742-4111}
\date{October 5, 2006}
\begin{abstract}
Carrier transport in gated 2D graphene monolayers is theoretically
considered in the presence of scattering by random charged impurity
centers with density $n_i$.  Excellent quantitative agreement is
obtained (for carrier density $n > 10^{12} \ \rm{cm}^{-2}$) with
existing experimental data (Ref. \onlinecite{kn:novoselov2004,
kn:novoselov2005, kn:zhang2005, kn:kim2006, kn:fuhrer2006}).  The
conductivity scales linearly with $n/n_i$ in the theory, and shows
extremely weak temperature dependence.  The experimentally observed
asymmetry between electron and hole conductivities is explained by the
asymmetry in the charged impurity configuration in the presence of the
gate voltage, while the high-density saturation of conductivity 
for the highest
mobility samples is explained as a crossover between the long-range
and the point scattering dominated regimes.  We argue that the
experimentally observed saturation of conductivity at low density
arises from the charged impurity induced inhomogeneity in the graphene
carrier density which becomes severe for $n \lesssim n_i \sim 10^{12} \
\rm{cm}^{-2}$.
\end{abstract}
\maketitle

Recent experimental observation of the density dependent 2D carrier
transport in single monolayers of graphene is an important, perhaps
seminal, development in low dimensional electronic phenomena in
nanostructures~\cite{kn:novoselov2004,kn:novoselov2005, kn:zhang2005,
kn:kim2006, kn:fuhrer2006}.  First, the gated 2D graphene systems
could potentially become multifunctional high speed and
high-power transistors, thus introducing a prospective paradigm shift
in the micro- (and nano-) electronics of the future.  Second, carriers
(both electrons and holes) have intriguing (and conceptually novel)
linear ``Dirac-like'' bare kinetic energy dispersion spectra in 2D
graphene monolayers due to the interesting honeycomb lattice structure
of Carbon atoms (with 2 atoms on inequivalent sublattices per unit
cell).  Graphene is, in fact, a carbon nanotube rolled out into a 2D
sheet, and the band structure induced carrier spectrum in graphene
monolayers is thus linear and chiral with a four-fold ground state
degeneracy arising from spin and valley, as well as a Berry phase term
arising from the inherent sublattice symmetry which also restricts the
carriers from back-scattering.  This theoretical absence of
back-scattering has led to the speculation that, as a matter of
principle, carrier mobilities in 2D graphene monolayers could be
extremely high even at room temperature, and indeed, already within
the first six months since the beginning of the subject (i.e. six
months since the fabrication of the first gated 2D graphene samples
with variable gate voltage tuned carrier density) mobilities as high
as $15000 \ \rm{cm}^{2}/V s$ have been reported from $1K$ to room
temperature.  These mobility values are comparable to (much higher than)
the mobility in the best available Si MOSFET samples at low (high)
temperatures.  It is, therefore, of considerable fundamental and
technological interest to investigate the theoretical limit on the 2D
graphene carrier mobility.

In this Letter, we present a theory for 2D graphene carrier transport
taking into account scattering by random charged impurity centers,
which is the most likely scattering mechanism limiting graphene
conductivity.  We also calculate the effect of short-range scattering
(e.g. by lattice defects), which may be present in some graphene
samples.  Quantitative agreement between our theory and existing
graphene experimental transport data is a strong indication that the
dominant carrier scattering mechanism operational in 2D
graphene monolayers is Coulomb scattering by random charged impurities
located in the substrate near the interface between the graphene layer
and the substrate.  We estimate the typical random charged impurity
(effective) concentration to be approximately $10^{12} \ \rm{cm}^{-2}$
in currently available graphene samples, and suggest that reducing
this impurity concentration to the $10^{10} \ \rm{cm}^{-2}$ density
range -- a difficult but not an impossible materials fabrications task
-- should increase 2D graphene mobility to the extremely large value
of $\sim 1.5 \rm{x} 10^{6} \ \rm{cm}^{2}/V s$ even at high
temperatures.  We argue that the experimentally observed ``low
density'' ($\lesssim 10^{12} \ \rm{cm}^{-2}$) saturation of 2D
graphene conductivity (i.e. the observed ``minimum conductivity'' at
nominally zero gate voltage) also arises from the presence of these
charged impurities which lead invariably to large-scale density
inhomogeneities in the 2D carrier system with the 2D density
fluctuations being larger than the average density at low carrier
density (i.e. near zero gate voltage) where the system therefore
breaks into a random network of 2D electrons and hole conducting
puddles, producing a finite conductivity at zero gate voltage.  Around
zero gate voltage, the 2D graphene layer is thus not a simple
(homogeneous) zero-gap semiconductor (as intrinsic graphene is
expected to be), but a spatially inhomogeneous semi-metal with small
random puddles of electron and hole 2D liquids depending on the
details of the charged impurity configuration in the sample.

For our purpose, intrinsic graphene is essentially a zero-gap
semiconductor (with the Fermi level $E_F$ precisely at $E=0$), with
linear chiral carrier kinetic energy dispersion (arising from the
known bulk band structure calculations) given by $E = \hbar \gamma k$,
where $k$ is the 2D carrier wavevector and $\gamma (\equiv v_F)$ is
the constant (i.e. independent of carrier density) Fermi velocity.
The intrinsic situation (i.e. with zero gate voltage) has no free
carriers at $T=0$ as a matter of principle (but, as a matter of
practice the system breaks up into spatially inhomogeneous conducting
puddles of 2D electron and hole droplets due to the potential
fluctuations induced by the extrinsic random charged impurity centers
which are invariably present in any real 2D graphene sample).  The
application of an external gate voltage (positive or negative) leads
to free carriers (electrons or holes) in the system.
We start by assuming the system to be a homogeneous 2D carrier system
of electrons (or holes) with a carrier density ``n'' induced by the
external gate voltage $V_g$ with $n = \kappa_s V_g / (4 \pi t)$, where
$\kappa_s$ is the dielectric constant of the substrate and $t$ is the
substrate thickness (i.e. the distance of the gate from the graphene
layer).  The chemical potential (or, equivalently the Fermi energy
$E_F$) at $T=0$ is given by $E_F = \hbar \gamma k_F$, where the 2D
Fermi wavevector $k_F$ depends on the carrier density through $k_F =
(4 \pi n /g_s g_v)^{1/2}$ with $g_s$ ($g_v$) being the
spin (valley) degeneracy of graphene.  (We choose $g_s = g_v = 2$
throughout this manuscript, although it is possible in some special
situations for these degeneracies to be lifted.)  

For large carrier densities, where the system is homogeneous, we
expect the full Boltzmann transport theory developed below to be
exact, hence providing an accurate determination of the charge
impurity density $n_i$ by comparing our theory with experimental
conductivity data, since $n_i$ just sets the overall scale of the
conductivity in the theory.  This in turn provides an estimate for the
breakdown of our transport theory (typically around $n \lesssim n_i$)
and the onset of the inhomogeneous puddles of electron and hole
carriers.  The experiments of Ref.~\onlinecite{kn:novoselov2005} show
that the characteristic density for this breakdown (i.e. the onset of
inhomogeneity in our opinion) is $n_c \sim n_i \approx 10^{12} \
\rm{cm}^{-2}$.  An independent estimate (giving the same order of
magnitude) can be made by considering the typical local density
fluctuations induced by a single charge located inside the substrate
at a distance $d$ from the interface.  This density fluctuation
becomes global in the presence of multiple impurities which can be
seen by considering multiple charged impurities distributed in a
random Poisson manner in a 2D plane at a distance $d$ above the
graphene monolayer.  One finds that the density fluctuation $\delta n$
around the gate-induced average 2D carrier density $n$ is given by
$\delta n^2 = n_i/(8 \pi d^2)$ confirming that transport close to the
Dirac point is dominated by a spatially inhomogeneous carrier density.
The transport through such 2D puddles is essentially a one dimensional
random network for each of the four spin/valley channels providing a
conductivity $\sigma \sim e^2/h$, although a more complete theory is
necessary to provide any quantitative comparison with experiments for
$n \lesssim n_c \sim n_i$.

We now proceed to describe in detail the microscopic transport
properties at high carrier density using the Boltzmann transport
theory~\cite{kn:dassarma}.  We calculate the
mobility in the presence of randomly distributed Coulomb impurity
charges near the surface with the electron-impurity interaction being
screened by the 2D electron gas in the random phase approximation
(RPA). The screened Coulomb scattering is the only important
scattering mechanism in our calculation.  There are additional
interface-scattering mechanisms unrelated to the Coulomb centers
(e.g. surface roughness scattering), but such interface scattering is
believed to be less important in the case of graphene (see below).

We also neglect all phonon scattering effects mainly because both the
reduced phase space imposed by chirality and the low temperatures
cause phonon scattering to be negligible in the regime of current
interest.  Given that 2D graphene is essentially a weakly interacting
system with effective $r_s \sim 0.75$, a constant independent of
carrier density ($r_s$ here is just the effective fine structure
constant), we expect our Boltzmann theory to be a quantitatively and
qualitatively accurate description of graphene transport for all
practical purposes.

In Boltzmann theory the conductivity for graphene is given by $\sigma
= (e^2/h) (2 E_F \langle \tau \rangle /\hbar)$ which comes from the
massless chiral Dirac spectrum, where $\langle \tau \rangle$, the
energy averaged finite temperature scattering time, is given by

\begin{eqnarray}
\langle \tau \rangle = \frac{{\int d\epsilon_k \epsilon_k \tau(\epsilon_k)
  \left ( -\frac{\partial f}{\partial \epsilon_k}
\right )}}{{\int d\epsilon_k \epsilon_k 
\left ( - \frac{\partial f}{\partial \epsilon_k} \right )}},
\end{eqnarray}
where $f(\epsilon_k)$ is the Fermi distribution function, 
$f(\epsilon_k) =\{ 1+\exp[(\epsilon_k-\mu)]/k_B T \}^{-1}$ 
with $\mu(T,n)$ as the finite temperature chemical potential determined 
self-consistently.

The energy dependent scattering time $\tau(\epsilon_k)$ for our model of
randomly distributed impurity charge centers is given in the
leading-order theory by

\begin{eqnarray}
\label{eq:scattime}
\frac{1}{\tau(\epsilon_k)} & = & \frac{\pi}{\hbar}\sum_a  n_i^{(a)} 
\int\frac{d^2k'}{(2\pi)^2}
\left |\frac{v^{(a)}(q)}{\varepsilon(q)}\right |^2 \nonumber \\
& & \times (1-\cos\theta) (1 + \cos\theta) \delta\left (
\epsilon_{\bf k} - \epsilon_{\bf k'} \right ),
\end{eqnarray}
where $n_i^{(a)}$ is the concentration of the $a$-th kind of 
impurity center, $q = |{\bf k} - {\bf k}'|$, $\theta \equiv
\theta_{{\bf kk}'}$ is the scattering angle between the scattering in-
and out- wave vectors ${\bf k}$ and ${\bf k}'$, $\epsilon_{\bf k} =
\hbar \gamma |{\bf k}|$, and $v^{(a)}(q,d)$ is the matrix elements
of the scattering potential between an
electron and an impurity. For Coulomb interaction, we use
 $v^{(c)}(q,d) = 2 \pi e^2 \exp(-qd) /(\tilde{\kappa} q)$ where $d$ is
the location of the charge impurity in the substrate 
measured from the interface and for short-range point defect scatterers, 
$v^{(p)}(q,d) = v_0$, a constant.  In Eq.~\ref{eq:scattime},
$\varepsilon(q)$ is the 2D finite temperature static RPA
dielectric (screening) function appropriate for 
graphene~\cite{kn:Hwang2006b}, given by
\begin{widetext} 
\begin{eqnarray}
\varepsilon(q) = 1 + \frac{q_s}{q} \left\{ 
 \begin{array}{ll} 1 - \frac{\pi q}{8 k_F} & \mbox{if $q \leq 2 k_F$} \\
                  1 - \frac{1}{2}\sqrt{1- \left(\frac{2 k_F}{q}\right)^2}
		  - \frac{q}{4 k_F} \sin^{-1}\left(\frac{2 k_F}{q} \right)
                  & \mbox{if $q > 2 k_F$} \end{array} 
\right. , 
\end{eqnarray}
\end{widetext}
where $q_s = 4 e^2 k_F/(\hbar \tilde{\kappa} \gamma)$ is the effective
graphene 2D Thomas-Fermi wavevector and $\tilde{\kappa} = \kappa(1 +
\pi r_s/2)$ is the effective dielectric constant.  In calculating the
conductivity in Eq.~\ref{eq:scattime}, the factor $(1+\cos\theta)/2$
arises from the sublattice symmetry (related to Berry phase) which
suppresses the backscattering contribution to
resistivity.  We can immediately observe from
Eq.~\ref{eq:scattime}, that $\tau^{-1} = \tau_c^{-1} + \tau_s^{-1}$,
where the Coulomb scattering time $\tau_c \sim \sqrt{n}$ dominates at
low density over the short range scattering time $\tau_s \sim
1/\sqrt{n}$. 

\begin{figure}
\bigskip
\epsfxsize=0.8\hsize
\hspace{0.0\hsize}
\epsffile{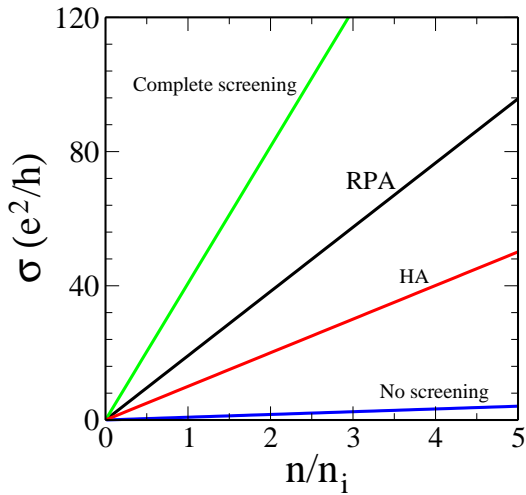}
\caption{\label{Fig:RPAvsTF} Graphene conductivity 
limited by Coulomb scattering calculated using different approximation
schemes with dielectric constant $\kappa = 2.5$.  The complete
screening approximation (Ref.\protect{~\onlinecite{kn:nomura2007}})
does not depend on dielectric constant and overestimates the
conductivity, whereas using the unscreened dielectric function (also
in Ref.\protect{~\onlinecite{kn:nomura2007}}) would have conductivity
less than $4e^2/h$ for the entire range of gate voltages used in the
experiment.  RPA is the main approximation that we use in this work,
and HA is the Hubbard approximation which incorporates local field
corrections.}
\end{figure}

We emphasize that in order to get quantitative agreement with
experiment, the full RPA dielectric function needs to be
calculated. In Fig.~\ref{Fig:RPAvsTF} we show comparison between
calculated conductivities in the RPA screening used in our calculation
with other approximation schemes including the completely screened 
(``strong screening'') approximation~\cite{kn:nomura2007} valid
only for $r_s \gg 1$, which gives a larger value for the
conductivity that is independent of the background lattice dielectric
constant $\kappa$.  We note that for graphene $r_s \sim 0.75$ making
the strong screening approximation invalid.  We also show results for
Hubbard approximation (HA) screening which includes local field
corrections approximately.  A recent
theory~\cite{kn:cheianov2006} has considered the temperature
dependence of conductivity within the Thomas-Fermi approximation.

\begin{figure}
\bigskip
\epsfxsize=0.8\hsize
\hspace{0.0\hsize}
\epsffile{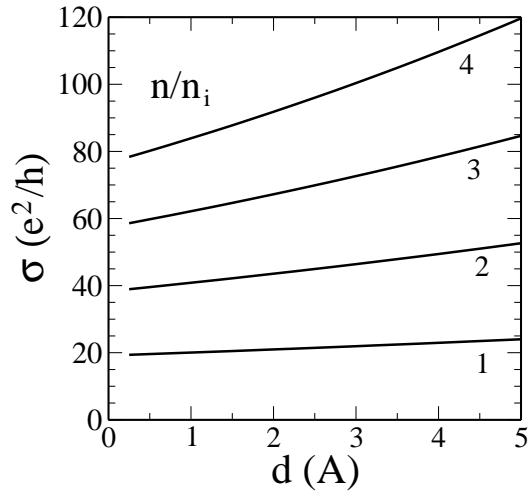}
\caption{\label{Fig:CompDist} The effect of remote scatters.
Here $d$ is the distance between the 2D graphene layer and
the 2D impurity layer.}
\end{figure}

Our main result is the quantitative agreement with
experiments in the regime where the conductivity is linear in density.
We note, however, that electron-hole asymmetry with a super-linear
conductivity is ubiquitous in the experimental data.  Changing the
sign of bias voltage has no effect on the concentration of charged
impurities, but (and especially at high voltages) it changes the
average distance between 2D graphene sheet and the impurity in the
substrate, particularly since the random charged impurities could be
both positive and negative in unequal numbers.  In our model, we
approximate that the impurities are confined in a 2D plane located at
a distance $d$ from the interface.  We find that a small shift of $d
\sim 1-2$~\AA \ is sufficient to explain the super-linear behavior
seen in the experiments.  As shown in Fig.~\ref{Fig:CompDist}, we
notethat for fixed $d$, the difference between the actual conductivity
and the linear value at $d=0$ is negligible at low density, $n/n_i
\sim 1$, but becomes important for $n/n_i \gtrsim 3$.

\begin{figure}
\bigskip
\epsfxsize=0.8\hsize
\hspace{0.0\hsize}
\epsffile{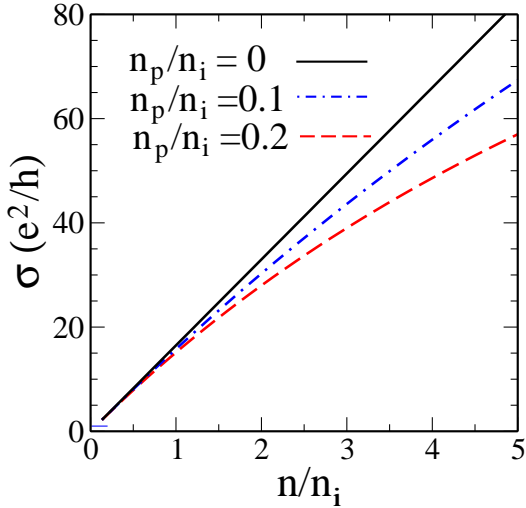}
\caption{\label{Fig:ShortAndLong} Graphene conductivity calculated using
a combination of short and long range scatterers.  One finds that
sub-linear conductivity at high density is likely to be seen in samples
with a small Coulomb impurity density and high mobility.}
\end{figure}

Another feature of the experiments is that for very high mobility samples, 
one finds a sub-linear conductivity~\cite{kn:kim2006}.  Such 
high quality samples presumably
have a small charge impurity concentration $n_i$ and 
it is therefore likely that  point defects here play a more dominant role.  
Point defects (which would be dominant for either 
large carrier density $n$ or low charge impurity concentration $n_i$) 
gives rise to a constant conductivity in contrast to charged
impurity scattering which produces a conductivity roughly linear
in $n/n_i$.  Any realistic graphene sample is 
somewhere in the crossover between these two limits.  Our formalism 
(see Eq.~\ref{eq:scattime}) can include both effects, where zero
range scatterers are treated with an effective point defect density of 
$n_p$.

We estimate that for most samples, we have $n_p/n_i \ll 1$, while for
the highest mobility samples, $n_p/n_i \sim 0.2$.  Shown in
Fig.~\ref{Fig:ShortAndLong} is the graphene conductivity calculated
including both charge impurity ($n_i$) and zero range point defect
($n_p$) scattering for different ratios of the point scatterer
impurity density $n_p$ and the charge impurity density $n_p$.  For
small $n_p/n_i$ we find the linear conductivity that is seen in most
experiments.  For large $n_p/n_i$ we see the flattening out of the
conductivity curve (which in the literature~\cite{kn:zhang2005} been
referred to as the sub-linear conductivity). We believe this
high-density flattening of the graphene conductivity~\cite{kn:kim2006}
is a non-universal crossover behavior arising from the competition
between two kinds of scatterers, $n_p$ and $n_i$.

\begin{figure}
\bigskip
\epsfxsize=1\hsize
\hspace{0.0\hsize}
\epsffile{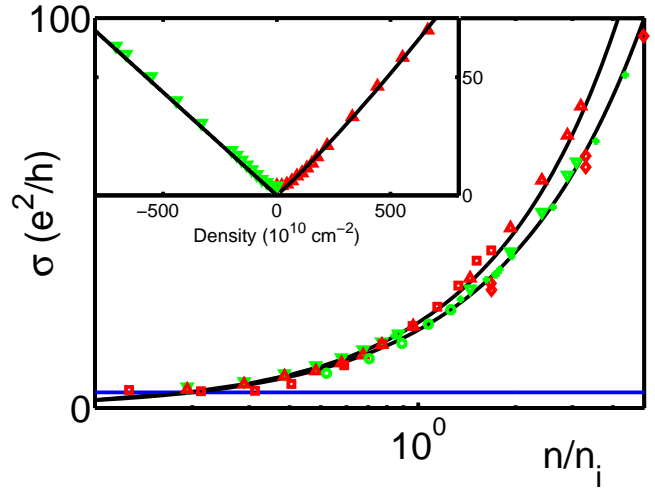}
\caption{\label{Fig:CompExpt} Figure shows
comparison of theory (Eq.~\ref{eq:scattime}) with experimental data
of Ref.~\onlinecite{kn:novoselov2005} (up and down triangles, 
$n_i = 2.3 \rm{x} 10^{12} cm^{-2}$),  Ref.~\onlinecite{kn:fuhrer2006} 
(circles and squares, $n_i = 3.4 \rm{x}  10^{12} cm^{-2}$) and 
Ref.~\onlinecite{kn:kim2006} 
(diamonds and crosses, $n_i = 0.43 \rm{x} 10^{12} cm^{-2}$) for electrons
and holes respectively. Solid lines (from bottom to top) 
show the minimum conductivity of $ 4 e^2/h$, theory for d=0, 
and $d=2$~\AA.  The inset shows Ref.~\onlinecite{kn:novoselov2005} data 
on a linear scale
assuming that the impurity
shifts by $d=2$~\AA~ for positive voltage bias.}
\end{figure}

Finally, in Fig.~\ref{Fig:CompExpt} we show the comparison of 
our theory (with only charged impurity scattering) to 
recent experiments~\cite{kn:novoselov2005, kn:kim2006, kn:fuhrer2006}.  
We determine $n_i$ (which simply sets the overall scale) 
by fitting the data at high density to the 
result of Eq.~\ref{eq:scattime} and observe very good
agreement between all three experimental data sets 
and our transport theory.  
We note that the order of magnitude for $n_i \sim 10^{12} \
\rm{cm}^{-2}$ agrees with independent estimates (e.g. the low-density
$n \lesssim n_i$ conductivity saturation around zero gate voltage).
The one-parameter (i.e. $n_i$) quantitative agreement between our theory
and three independent sets of experimental data is one of our 
most important findings.   

We emphasize that our Drude-Boltzmann semiclassical transport theory
is valid only in the high density, $n > n_i$, regime and therefore
only the agreement between theory and experiment in the high density
regime is meaningful.  Indeed, our theory predicts that conductivity
should vanish linearly as carrier density goes to zero, whereas in
reality, the graphene conductivity at low density (i.e. gate voltage),
$n < n_i$ in our model, becomes approximately a density independent
constant of the order $e^2/h$.  Our theory provides no quantitative or
qualitative explanation for this low-density behavior, where we
believe that the physics of strong density inhomogeneity (random
puddles of 2D electron and holes) induced by the charged impurities in
the insulating dielectric dominates, and our Drude-Boltzmann physics
simply does not apply.  We cannot, however, rule out that the
low-density regime is dominated by localization physics
~\cite{kn:fradkin}, or the Dirac
cone physics of intrinsic graphene~\cite{kn:ludwig} since 
our semiclassical theory does not apply in the
$n < n_i$ low density regime.  We note, however, that recent numerical
work~\cite{kn:nomura2007, kn:rycerz2006} on quantum transport in the
low density regime is consistent with the picture of a percolation
transition between electron and hole carriers with the implication
that the zero-carrier density intrinsic regime may simply be
experimentally inaccessible.  However, what we have convincingly
established in this work is that the high-density, $n > n_i$, $2D$
graphene transport is dominated by impurity scattering which can be
theoretically described by a microscopic Drude-Boltzmann model.    
 
In conclusion we have developed a detailed microscopic transport
theory for 2D graphene layers.  We have shown that charged impurities
in the substrate are the dominant source of scattering.
We have argued that the low density regime is dominated by density fluctuations
caused by these charge impurities and that a picture of inhomogeneous puddles
of conducting electrons and conducting holes is necessary to
understand the finite value of conductivity near zero bias.  Finally
the Fermi temperature of 2D graphene being around $1300 K$ for $n \sim
10^{12} \ \rm{cm}^{-2}$, there is essentially no temperature
dependence in 2D graphene conductivity in the $0-300K$ temperature
range arising from charged impurity scattering, a fact that is in
excellent agreement with experimental observation~\cite{kn:geim2006}.    

We would like to thank M. Fuhrer, A. Geim, and P. Kim for sharing 
with us their unpublished data, and V. Galitski and W-K. Tse for 
discussions. This work was partially supported by U.S. ONR.

\end{document}